\documentclass[twocolumn,twocolappendix]{aastex631}
\usepackage[cjk]{kotex}

\received{}
\revised{}
\accepted{}
\submitjournal{ApJ}

\shorttitle{}
\shortauthors{Yu}
\begin{document}
\small
\title{Dissipative Instability of Magnetohydrodynamic Sausage Waves in a Compressional Cylindrical Plasma: Effect of Flow Shear and Viscosity Shear}

\correspondingauthor{D. J. Yu}
\email{djyu79@gmail.com}
\author[0000-0003-1459-3057]{D. J. Yu (유대중)}
\affiliation{Department of Astronomy and Space Science, Kyung Hee University,
 1732, Deogyeong-daero, Yongin, Gyeonggi 17104, Republic of Korea}

\begin{abstract}
The shear flow influences the stability of magnetohydrodynamic (MHD) waves. In the presence of a dissipation mechanism, flow shear may induce a MHD wave instability below the threshold of the Kelvin-Helmholtz instability (KHI), which is called dissipative instability (DI). This phenomenon is also called negative energy wave instability (NEWI) because it is closely related to the backward wave which has negative wave energy. Considering viscosity as a dissipation mechanism, we derive an analytical dispersion relation for the slow sausage modes in a straight cylinder with a discontinuous boundary. It is assumed that the steady flow is inside and dynamic and bulk viscosities are outside the circular flux tube under photospheric condition. When the two viscosities are weak, it is found that for the slow surface mode, the growth rate is proportional to the axial wavenumber and flow shear, consistent with in the incompressible limit. For a slow body mode, the growth rate has a peak at certain axial wavenumber and its order of magnitude is similar to surface mode. The linear relationship between the growth rate and the dynamic viscosity established in the incompressible limit develops nonlinearly when the flow shear {and/or} the two viscosities are sufficiently strong.

\end{abstract}

\keywords{magnetohydrodynamics (MHD) -- waves -- Sun: oscillations --  Sun: photosphere}

\section{Introduction} \label{sec:intro}

The non-uniform structure of solar atmosphere influences the behaviors of the magnetohydrodynamic (MHD) waves.
The shear flow substantially modifies the characteristics of the wave modes (e.g., Nakariakov \& Roberts~\citeyear{Nakariakov1995}, Yu \& Nakariakov~\citeyear{Yu2020}, Skirvin et al.~\citeyear{Skirvin2022}).
The Kelvin-Helmholtz instability (KHI) may arise due to the background shear flow  (Chandrasekhar~\citeyear{Chandrasekhar1961}, Rae~\citeyear{Rae1983}, Zhelyazkov \& Zaqarashvili~\citeyear{Zhelyazkov2012}), or due to the dynamic velocity shear induced by waves (Magyar \& Van Doorsselaere~\citeyear{Magyar2016}). The KHI is of particular importance for its cause of mixing or turbulence which leads to heating via energy transfer into smaller scales (Karpen et al.~\citeyear{Karpen1993}, Karampelas \& Van Doorsselaere~\citeyear{Karampelas2018}, Guo et al.~\citeyear{Guo2019}).

When there is a dissipative process, shear flow may induce another type of wave instability called dissipative instability (DI) where the viscosity shear is usually adopted for the dissipation mechanism (Cairns~\citeyear{Cairns1979}).
The DI is closely connected to the occurrence of the negative energy wave (NEW). The backward wave has negative energy when the flow shear exceeds a threshold called critical speed, which is lower than the threshold for KHI, and copropagates with the forward wave (Cairns~\citeyear{Cairns1979}, Joarder et al.~\citeyear{Joarder1997}, Ruderman \& Goossens~\citeyear{Ruderman1995}, Yu \& Nakariakov~\citeyear{Yu2020}).

Studies for the DI range from the convection zone to the corona.
Ryutova~(\citeyear{Ryutova1988}) first addressed the importance of the negative energy wave (NEW) in the solar atmosphere. She studied the kink modes in the long wavelength limit in the thin magnetic flux tube.
Ruderman \& Goossens (\citeyear{Ruderman1995}) considered the propagating surface Alfv\'{e}n wave with negative energy in a plasma slab, showing its increment is in proportion to the viscosity. Anisotropic viscosity and thermal conductivity in a compressible plasma was considered  by Ruderman et al.~(\citeyear{Ruderman1996}), who showed that the presence of viscosity lowers the threshold for instability.
Joarder et al.~(\citeyear{Joarder1997}) discussed the criterion for the NEW in a plasma slab structure where photosphere, corona, and solar wind were of interest.
Holzwarth et al.~(\citeyear{Holzwarth2007}) investigated the Parker-like instability (PI), KHI, and DI in the convection zone, obtaining that the critical speed is below the thresholds for the KHI and PI.
Considering the radiative loss as a dissipation mechanism for the slow body modes in a coronal jet, Pourjavadi et al.~(\citeyear{Pourjavadi2021}) found that DI for slow sausage modes is weaker than for slow kink modes except where the axial wavelength goes to infinity.
The DI was shown to become more complicate by considering ionization degree for the prominence plasma slab ({Ballai et al.~\citeyear{Ballai2015}}, Ballai et al.~\citeyear{Ballai2017}).
Yu \& Nakariakov (\citeyear{Yu2020}) studied the DI of MHD surface modes in an incompressible plasma cylinder, pointing out that the sausage mode
($m=0$) and the higher modes ($m\ne0$) have different behaviors.

{DI can occur in a similar situation for KHI since a damping (decay) mechanism is added to the situation of KHI.}
Despite the efforts for some decades, the DI is not well-understood yet {and there is still no analytical theory for DI in a compressional plasma cylinder}.
{The primary aim of this paper is to develop an analytical theory of the DI in a compressional plasma, generalizing the theory of Yu \& Nakariakov (\citeyear{Yu2020}).
As a first step, we focus on the slow sausage modes under photospheric condition, where DI is induced by viscosity shear and flow shear in a cylindrical compressional plasma penetrated by an axial magnetic field. In the photosphere the flow speed is in the range of sound speed therein (e.g., Keil et al. \citeyear{Keil1999}), thus the slow modes can be of concern for DI.}

The paper is organized as follows.
We describe the model and develop the analytical theory in Sec.~\ref{sec:model}. The numerical results are shown in Sec.~\ref{sec:results} and conclusions in Sec.~\ref{sec:conclusion}.

\section{Model} \label{sec:model}
The governing equations are the viscous MHD equations for a compressible plasma (e.g., Priest \& Forbes \citeyear{Priest2007}):
\begin{eqnarray}
\frac{\partial \rho}{\partial t}+\nabla\cdot(\rho\textbf{\textsl{v}})&=&0,\label{eq:1}\\
\rho\frac{\partial \textbf{\textsl{v}}}{\partial t}+\rho\textbf{\textsl{v}}\cdot\nabla\textbf{\textsl{v}}+\nabla p-\textbf{\textsl{j}}\times\textbf{\textsl{B}}&=&\mu\nabla^2\textbf{\textsl{v}}+\varsigma\nabla(\nabla\cdot\textbf{\textsl{v}}),~~
\label{eq:2}\\
\frac{\partial p}{\partial t}+\textbf{\textsl{v}}\cdot\nabla p+\gamma_0 p\nabla\cdot\textbf{\textsl{v}}&=&0,\label{eq:3}\\
\frac{\partial \textbf{\textsl{B}}}{\partial t}-\nabla\times(\textbf{\textsl{v}}\times\textbf{\textsl{B}})&=&0,\label{eq:4}\\
\textbf{\textsl{j}}-\frac{1}{\mu_0}\nabla\times\textbf{\textsl{B}}&=&0,\label{eq:5}
\end{eqnarray}
where $\rho$ is the density, $\textbf{\textsl{v}}$ is the velocity, $\textbf{\textsl{B}}$ the magnetic field, $p$ the pressure, and $\textbf{\textsl{j}}$ the electric current density, $\mu_0$ is the permeability of vacuum and $\gamma_0(=5/3)$ is the ratio of specific heat, $\varsigma=\zeta+\mu/3$, $\mu$ is the shear (dynamic) viscosity, and $\zeta$ is the bulk viscosity. {Here we ignore the effects of gravitation and partial ionization.}

We consider a cylindrical flux tube with the radius $R$, of which the inside and outside regions are homogeneous with different parameter values. It is assumed that the equilibrium maintains the pressure balance
\begin{eqnarray}
p_i+\frac{B_i^2}{2\mu_0}&=&p_e+\frac{B_e^2}{2\mu_0},\label{eq:6}
\end{eqnarray}
where the subscript $i(e)$ denotes the parameter inside (outside) the flux tube.
The background axial flow is assumed to be in the inside and the two viscosities in the outside.
From the viscous MHD equations, we derive linear wave equations for each regions by applying Fourier transform with the factor $\exp[-i(\omega t-k_zz)]$ where $\omega$ is the angular frequency of the wave and $k_z$ is the axial wavenumber. We then obtain the dispersion relation with the help of the matching conditions at the flux tube boundary ($r=R$). We also obtain the criterion and critical speed for the negative energy wave.

\subsection{Wave equation inside the flux tube}\label{sec:2-1}
When there are steady background flow ($\textbf{\textsl{U}}_0=(0,0,U)$) and magnetic field ($\textbf{\textsl{B}}_0=(0,0,B_i)$) in the axial direction, we have two coupled differential equations for the perturbed total pressure $P$ and the radial Lagrangian displacement $\xi_r$ (e.g., Goossens et al.~\citeyear{Goossens1992}) as
\begin{eqnarray}
 D_1\frac{d(r\xi_{ri})}{dr}&=&-rC_{1}P_i,\label{eq:7}\\
D_1\frac{dP_i}{dr} &=&C_{2}\xi_{ri},\label{eq:8}
\end{eqnarray}
where
\begin{eqnarray}
D_1&=& \rho_i(v_{si}^2+v_{Ai}^2)(\Omega^2-\omega_{Ai}^2)(\Omega^2-\omega_{ci}^2),\label{eq:9} \\
C_{1}&=& \Omega^4-k_z^2(v_{si}^2+v_{Ai}^2)(\Omega^2-\omega_{ci}^2),\label{eq:10}\\
C_{2}&=&\rho_i^2(v_{si}^2+v_{Ai}^2)(\Omega^2-\omega_{Ai}^2)^2(\Omega^2-\omega_{ci}^2),\label{eq:11}\\
\Omega &=&\omega-k_zU,\label{eq:12}
\end{eqnarray}
and $\omega_{Ai}=v_{Ai}k_z$, $\omega_{ci}=v_{ci}k_z$, $v_{Ai}=B_i/\sqrt{\rho_i\mu_0}$ is Alfv\'{e}n speed, $v_{si}=\sqrt{\gamma_0 p_i/\rho_i}$ is the sound speed and $v_{ci}=v_{si}v_{Ai}/\sqrt{v_{si}^2+v_{Ai}^2}$. The radial displacement $\xi_{ri}$ has the relation with the perturbed radial velocity $v_{ri}$ (Goossens et al.~\citeyear{Goossens1992})
\begin{eqnarray}
\xi_{ri} &=&\frac{i}{\Omega}v_{ri}.\label{eq:13}
\end{eqnarray}
Eqs.~(\ref{eq:7}) and (\ref{eq:8}) have Bessel functions as solutions.

\subsection{Wave equation outside the flux tube}\label{sec:2-2}
For the outside of the flux tube we assume viscosity terms and no background flow where $\textbf{\textsl{B}}_0=(0,0,B_e)$.
Instead of deriving wave equations for the displacement $\xi$ and the perturbed total pressure $P$, we derive a governing wave equation for the divergence of the perturbed velocity $\triangle\equiv\nabla\cdot\textbf{v}_1=\psi(r)\exp[-i(\omega t-k_zz)]$ (see Appendix~\ref{sec:apped1}):
\begin{eqnarray}\label{eq:14}
\omega^4\triangle_e&+&\bigg[(\bar{v}_{se}^2+v_{Ae}^2)\omega^2-\bigg(v_{Ae}^2\bar{v}_{se}^2k_z^2+\frac{i\mu}{\rho_e}\omega^3\bigg)
\bigg]\nabla^2{\triangle_e}\nonumber\\
&-&\frac{i\mu\omega}{\rho_e}(\bar{v}_{se}^2+v_{Ae}^2)\nabla^4{\triangle_e}=0,
\end{eqnarray}
where $\nabla^2$ is Laplacian and $\bar{v}_{{se}}^2=v_{se}^2-i\omega\frac{\mu+\varsigma}{\rho_e}$.
This equation can be further reduced to
\begin{eqnarray}\label{eq:15}
&&A\bigg[\frac{d^4\psi(r)}{dr^4}+\frac{2}{r}\frac{d^3\psi(r)}{dr^3}\bigg]
-\bigg[A\bigg(2k_z^2+\frac{1}{r^2}\bigg)-1\bigg]
\frac{d^2\psi(r)}{dr^2}\nonumber\\
&&~~-\bigg[A\bigg(2k_z^2-\frac{1}{r^2}\bigg)-1\bigg]\frac{1}{r}\frac{d\psi(r)}{dr}
-n^2\psi(r)=0,
\end{eqnarray}
where
\begin{eqnarray}
A&=&\frac{i\mu\omega}{\rho_e(\bar{\omega}_{ce}^2-\omega^2)},\label{eq:16}\\
n^2&=&\frac{\omega^4-k_z^2(\bar{v}_{se}^2+v_{Ae}^2)\big(\omega^2-\bar{\omega}_{ce}^2+\frac{i\mu\omega}{\rho_e}k_z^2\big)}
{(\bar{v}_{se}^2+v_{Ae}^2)(\bar{\omega}_{ce}^2-\omega^2)},\label{eq:17}\\
\bar{\omega}_{ce}^2&=&\frac{\big(\bar{v}_{se}^2v_{Ae}^2k_z^2+\frac{i\mu}{\rho_e}\omega^3\big)}{(\bar{v}_{se}^2+v_{Ae}^2)}\label{eq:18}.
\end{eqnarray}
Eq.~(\ref{eq:15}) has Bessel functions of order zero as analytical solutions (e.g., Geeraerts et al.~\citeyear{Geeraerts2020}). In the present paper we use for the solution the modified Bessel function of second kind, $K_0(kr)$, where
\begin{eqnarray}
k=k_\pm=\sqrt{\frac{-1+2Ak_z^2\pm\sqrt{(1-2Ak_z^2)^2+4An^2}}{2A}}.\label{eq:19}
\end{eqnarray}
When $\mu(\varsigma)=0$, Eq.~(\ref{eq:15}) reduces to the well-known Bessel's differential equation.

As $k_\pm$ includes imaginary terms, the solution itself is a complex function. It was illustrated by Geeraerts et al.~(\citeyear{Geeraerts2020}) that the solution function behaves like body-like or surface-like mode depending on the values of the complex eigenfrequency obtained from the dispersion relation.

\subsection{Dispersion relation for the case with shear flow}\label{sec:2-3}
We first consider the situation with no viscosity. We assume
the background flow is inside the flux tube as in Sec.~\ref{sec:2-1} and no flow for the outside region. Using
the matching conditions at $r=R$ (e.g., Goossens et al.~\citeyear{Goossens1992}, Yu et al.~\citeyear{Yu2017a}, Sadeghi et al.~\citeyear{Sadeghi2021})
\begin{eqnarray}
\xi_{ri}&=&\xi_{re},\label{eq:20}\\
P_i&=&P_e,\label{eq:21}
\end{eqnarray}
we derive the dispersion relation as
\begin{eqnarray}
\rho_i(\Omega^2-\omega_{Ai}^2)-\rho_e(\omega^2-\omega_{Ae}^2)\frac{k_i}{k_e}Q_0&=&0,\label{eq:22}
\end{eqnarray}
where
\begin{eqnarray}
  k_i^2 &=& f\frac{(\Omega^2-\omega_{si}^2)(\Omega^2-\omega_{Ai}^2)}{(v_{si}^2+v_{Ai}^2)(\Omega^2-\omega_{ci}^2)},\label{eq:23} \\
  k_e^2 &=& -\frac{(\omega^2-\omega_{se}^2)(\omega^2-\omega_{Ae}^2)}{(v_{se}^2+v_{Ae}^2)(\omega^2-\omega_{ce}^2)},\label{eq:24}\\
  Q_0 &=&F_0(k_iR)\frac{K_0(k_eR)}{K_0'(k_eR)},\label{eq:25}
\end{eqnarray}
and
\begin{eqnarray}
f&=&\left\{ \begin{array}{ll}
     -1 & \mbox{ for surface modes}\\
      1 & \mbox{ for body modes }\end{array},~~~~~
         \right.\label{eq:26}\\
F_0(k_iR)&=&\left\{ \begin{array}{ll}
     \frac{I_0'(k_iR)}{I_0(k_iR)} & \mbox{ for surface modes}\\
      \frac{J_0'(k_iR)}{J_0(k_iR)} & \mbox{ for body modes }\end{array}.~~~~~
         \right.\label{eq:27}
\end{eqnarray}
Here $J_0(I_0)$ is the (modified) Bessel function of first kind and the prime denotes the derivative with respect to the entire argument.

\subsection{Negative energy waves}\label{sec:2-4}
The threshold of flow shear for NEW is called the critical speed.
When the flow shear is higher than the critical speed, the wave energy of the backward wave becomes negative. The criterion for the NEW is given as  (Cairns~\citeyear{Cairns1979}, Joarder et al.~\citeyear{Joarder1997}, Marcu~\citeyear{Marcu2007}, Yu \& Nakariakov~\citeyear{Yu2020})
\begin{eqnarray}
\mathcal{C}=\omega\frac{\partial \mathcal{D}}{\partial \omega}<0.\label{eq:28}
\end{eqnarray}
The dispersion function $\mathcal{D}$ has to be suitably defined for the characteristic function $\mathcal{C}$ to be positive when $U=0$.
We may define $\mathcal{D}$ as
\begin{eqnarray}
\mathcal{D}=\chi(\tilde{\Omega}^2-\tilde{\omega}_{Ai}^2)-(\tilde{\omega}^2-\tilde{\omega}_{Ae}^2)\frac{k_i}{k_e}Q_0,\label{eq:29}
\end{eqnarray}
where $\chi=\rho_i/\rho_e$ and tilde means the parameters are normalised by $\omega_{si}$. Similarly, the characteristic function $\mathcal{C}$ can be defined as $\mathcal{C}=v_p\frac{\partial\mathcal{D}}{\partial v_p}$ where $v_p=\omega/k_z$. We note that an analytical expression of $\mathcal{C}$ is possible when following the procedure to obtain $\frac{\partial \mathcal{D}}{\partial\omega}$ in Yu et al.~(\citeyear{Yu2017a}) or in Sadeghi et al.~(\citeyear{Sadeghi2021}).

The phase speed of the backward wave changes its sign when the flow shear crosses the critical speed. Therefore, one may obtain the critical speed by setting $\omega=0$ in the dispersion relation (\ref{eq:22}) (e.g., Yu \& Nakariakov~\citeyear{Yu2020}). See Fig.~\ref{fig:f5}.

\begin{figure}
\includegraphics[width=0.5\textwidth]{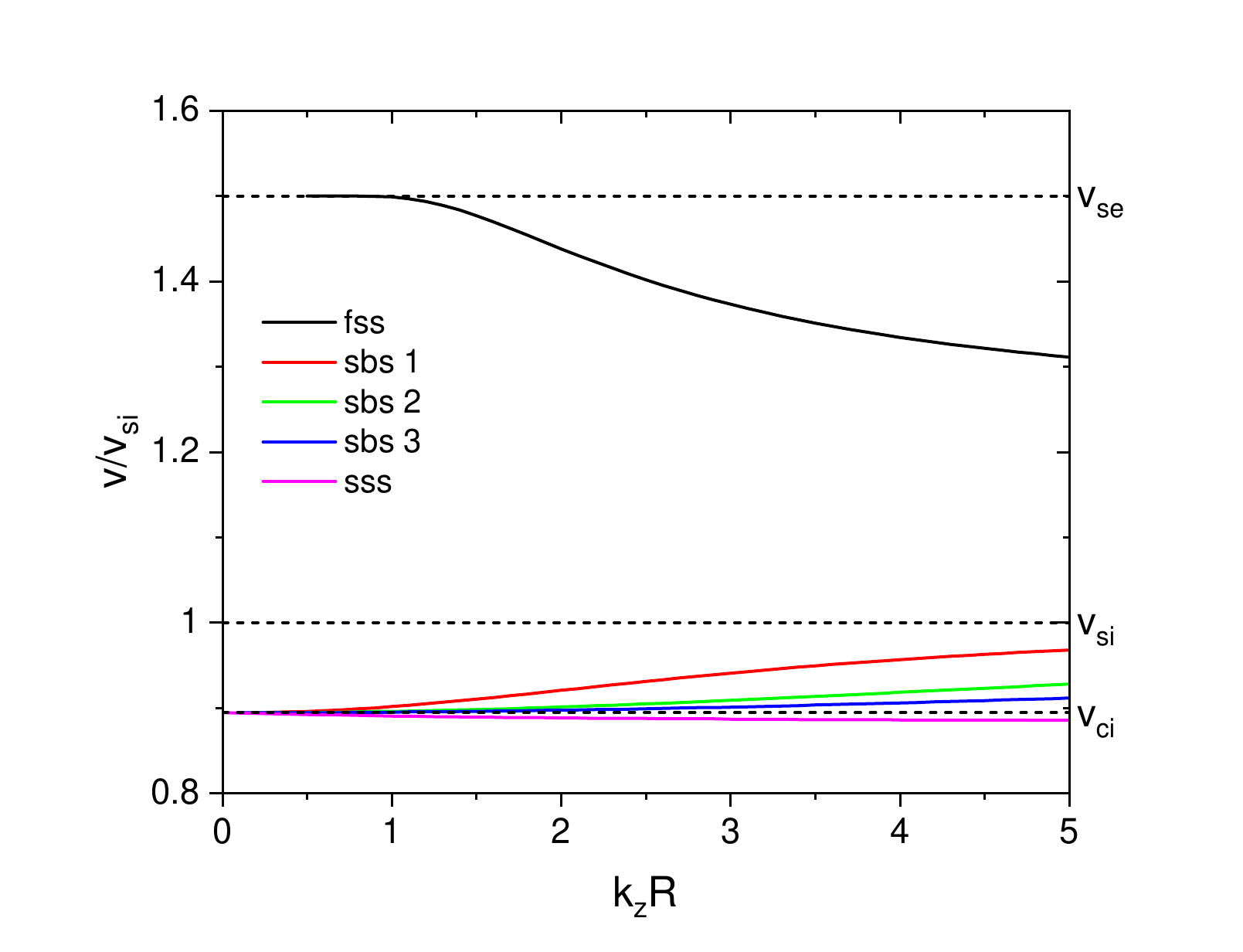}
\caption{\label{fig:f1} Phase speeds $v_+$ vs. $k_zR$ for the forward sausage modes under photospheric conditions where $v_{se}=1.5v_{si}$, $v_{Ai}=2.0v_{si}$, $v_{Ae}=0.5v_{si}$, $v_{ci}\approx0.8944v_{si}$ and $v_{ce}\approx0.4743v_{si}$: fast surface sausage mode (fss), slow body sausage modes (sbs 1-3), and slow surface sausage mode (sss).  These parameter values are used throughout the figures.}
\end{figure}

\begin{figure}
\includegraphics[width=0.42\textwidth]{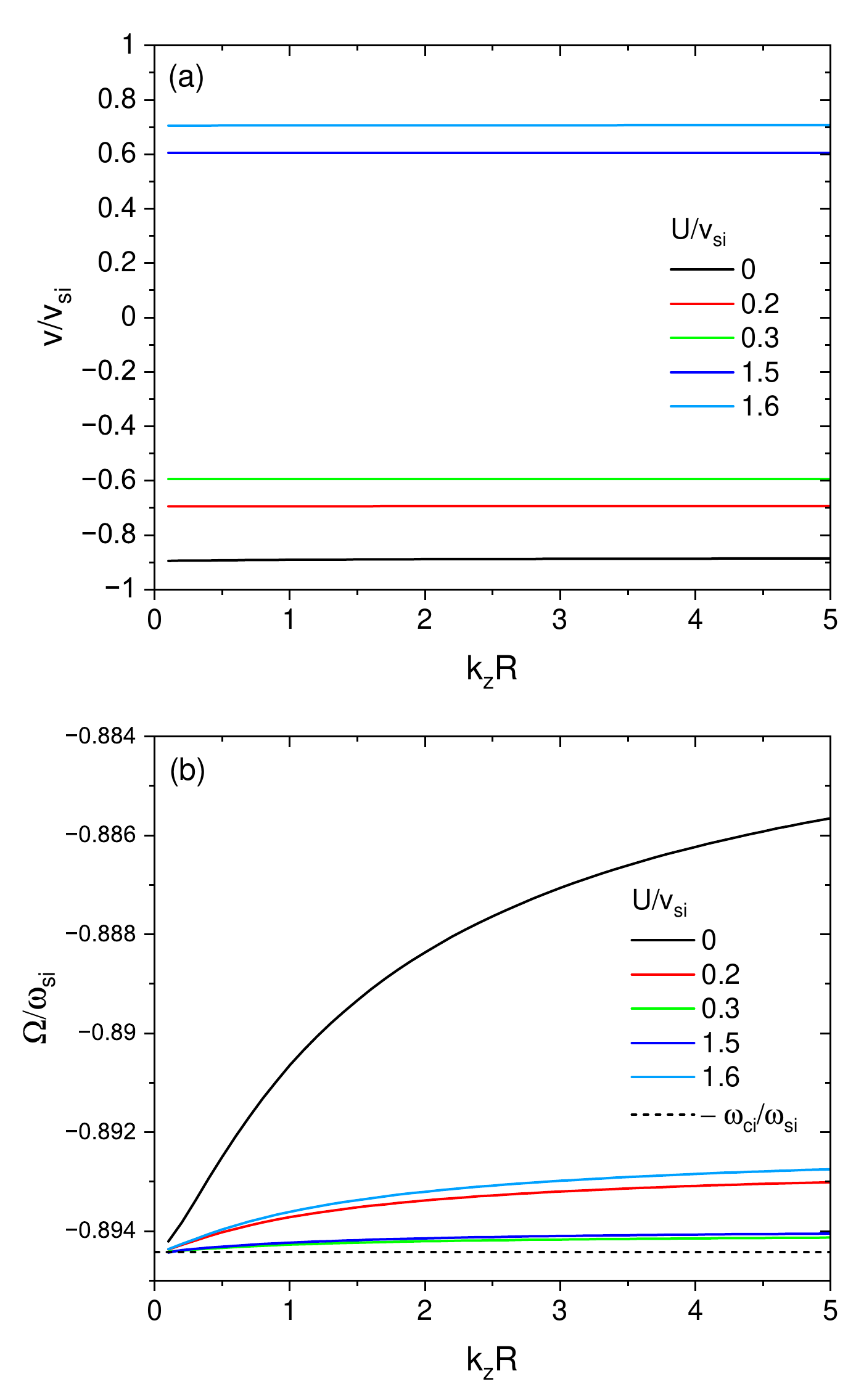}
\caption{\label{fig:f2} (a) Phase speed $v_-$ and (b) Doppler-shifted phase speed $\Omega_-$ of the backward slow surface sausage mode vs. $k_zR$. }
\end{figure}

\begin{figure}
\includegraphics[width=0.42\textwidth]{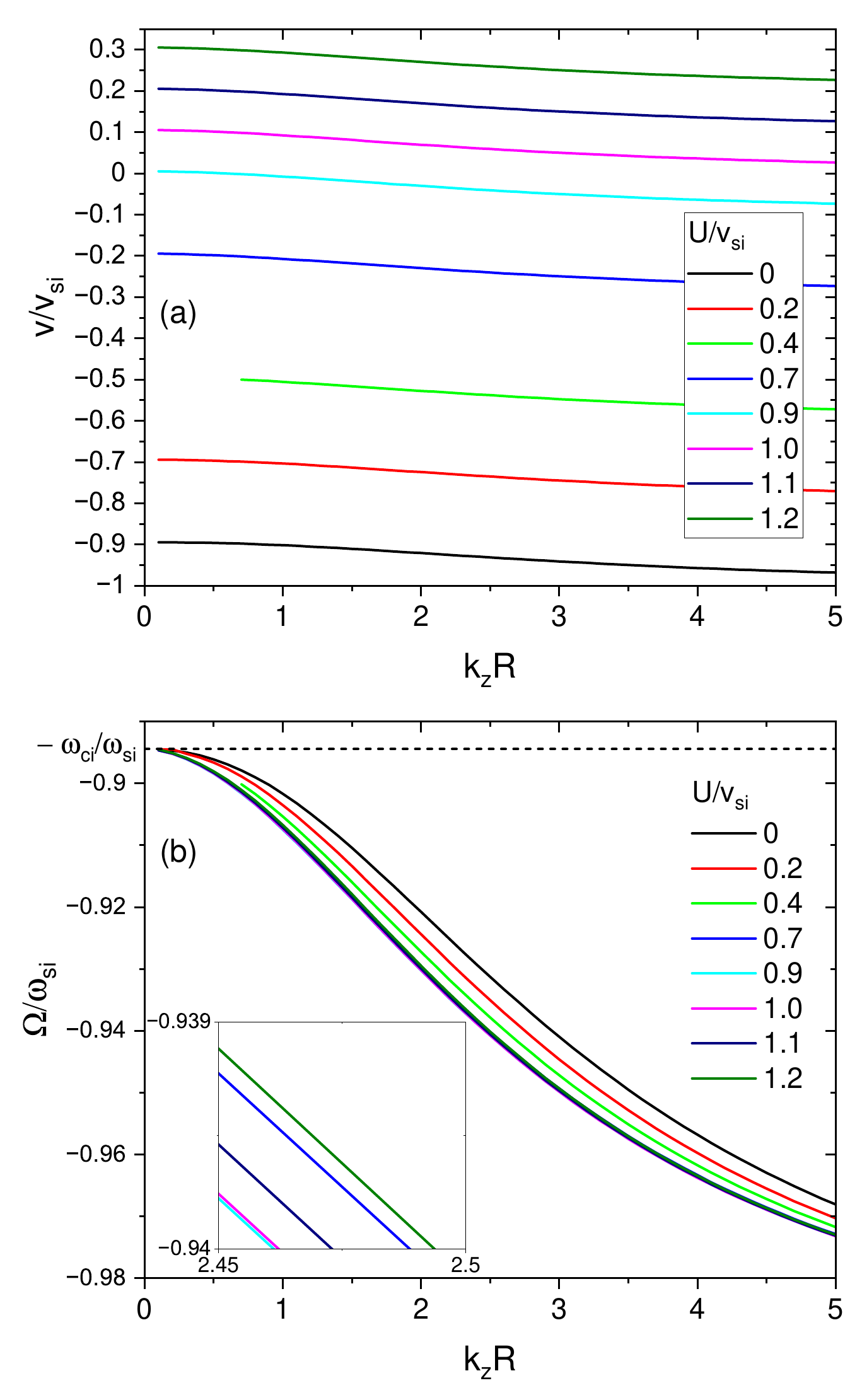}
\caption{\label{fig:f3} (a) $v_-$ and (b) $\Omega_-$ of the backward slow body sausage mode 1 (sbs 1) vs. $k_zR$. Inset in (b) illustrates the turning behavior of dispersion curve at $U/v_{si}\approx0.9$}
\end{figure}

\subsection{Dispersion relation with shear flow and viscosity}
Now we consider the situation that there exist flow inside and viscosity outside the flux tube.
When the viscosity terms are included there can exist two wave solutions outside the flux tube, so the wave solution $\psi(r)$ has in general the following form
\begin{eqnarray}
\psi(r)=\left\{ \begin{array}{ll}
     A_1\mathcal{F}_0(k_ir) & \mbox{ for $r< R $}\\
      A_2K_0(k_+r) +A_3K_0(k_-r)& \mbox{ for $ r> R$ }\end{array},\label{eq:30}
         \right.
\end{eqnarray}
where $A_1$, $A_2$, and $A_3$ are arbitrary complex constants, and
 \begin{eqnarray}
\mathcal{F}_0(k_ir)&=&\left\{ \begin{array}{ll}
     I_0(k_ir) & \mbox{ for surface modes}\\
     J_0(k_ir) & \mbox{ for body modes }\end{array}.~~~~~\label{eq:31}
         \right.
\end{eqnarray}
In the presence of viscosity outside the flux tube, the boundary condition (\ref{eq:21}) changes to
\begin{eqnarray}
P_i=P_e-2\mu\frac{\partial v_{re}}{\partial r}-\big(\zeta-\frac{2\mu}{3}\big)\triangle_e. \label{eq:32}
\end{eqnarray}
Since we have three unknown variables, $A_1$, $A_2$, and $A_3$, we need another boundary condition for the perturbed magnetic field in $z$ direction, $b_z$, (e.g., Geeraerts et al.~\citeyear{Geeraerts2022}):
\begin{eqnarray}
b_{zi}=b_{ze}. \label{eq:33}
\end{eqnarray}
Using Eqs.~(\ref{eq:20}), (\ref{eq:32}), and (\ref{eq:33}), we derive a matrix equation
\begin{eqnarray}
\pmatrix{ X_{1} & X_{2} & X_{3}
\cr X_{4} & X_{5} & X_{6} \cr
X_{7} & X_{8} & X_{9}}
\pmatrix{A_1 \cr A_2 \cr A_3}=\bf{X}\bf{A}=0
,\label{eq:34}
\end{eqnarray}
where $X_1,X_2,\ldots,X_9$ are shown in Appendix~\ref{sec:apped2}.

To have a non-trivial solution for Eq.~(\ref{eq:34}), It should be $\text{det}(\bf{X})=0$, which yields the dispersion relation
\begin{eqnarray}
&&X_1(X_5X_9-X_6X_8)-X_2(X_4X_9-X_6X_7)\nonumber\\&&+X_3(X_4X_8-X_5X_7)=0.\label{eq:35}
\end{eqnarray}
 Similarly to the solution for $\kappa_+$ in Geeraerts et al. (\citeyear{Geeraerts2020}), the solution for $k_-$ represents the electromagnetic boundary layer (Hartmann layer) at $r=R$. In the limit $\mu\rightarrow0$ and $\zeta\rightarrow0$, $|k_-|\rightarrow\infty$. The solution for $k_+$ corresponds to wave part.
For the present problem, the numerical results including the solution for $k_{-}$ are found to be physically unacceptable.
Thus, by ignoring the term for $k_-$ {($A_3=0$)}, we derive the following dispersion relation, using Eqs. (\ref{eq:20}) and (\ref{eq:32}),
\begin{eqnarray}
X_1X_5-X_2X_4=0.\label{eq:36}
\end{eqnarray}
{The use of Eq.~(\ref{eq:33}) for the boundary condition also gives wrong results.}
Eq.~(\ref{eq:36}) reduces to Eq. (\ref{eq:22}) when $\mu=\zeta=0$.


\section{Results} \label{sec:results}

\subsection{Dispersion curves with shear flow and without viscosity}\label{sec:3-1}
We first consider the behavior of the dispersion curves for the sausage modes depending on the flow speed $U$ in the absence of the viscosity.
The dispersion relation (\ref{eq:22}) allows two phase speeds $v_p=v_\pm$. The phase speed $v_+(v_-)$ corresponds to propagation in the positive (negative) $z$ direction with the relation $v_-=-v_+$ in the absence of the steady flow.
In Fig.~\ref{fig:f1} we plot the phase speed of the forward sausage modes, $v_+/v_{si}$, vs. $k_zR$ under a photospheric condition in the presence of no flow shear, where $v_{se}=1.5v_{si}$, $v_{Ai}=2.0v_{si}$, $v_{Ae}=0.5v_{si}$, $v_{ci}\approx0.8944v_{si}$ and $v_{ce}\approx0.4743v_{si}$ (Edwin \& Roberts~\citeyear{Edwin1983}). We use these parameter values through the paper. There appear three kinds of eigenmodes: fast surface (fss), slow surface (sss) and slow body (sbs) modes. Three body modes are plotted among multiple solutions.

Since the NEWI or DI occurs for the backward wave ($v_-$), we focus on the backward wave modes.
Fig.~\ref{fig:f2} (a) represents the dependence of the dispersion curve of the backward slow surface sausage (sss) mode, $v_-$, on the flow speed $U$. The phase speed $v_-$ tends to shift upward as $U$ increases, while there appears a gap when $0.3<U/v_{si}<1.5$. The Doppler-shifted phase speed in Fig.~\ref{fig:f2} (b) shows that in this range $\Omega<-\omega_{ci}$ and the slow surface mode is not allowed. When $\Omega>\omega_{ci}$ the slow surface mode also does not exist.

In Fig.~\ref{fig:f3} we plot $U$ dependence of (a) the phase speed $v_-$ and (b) Doppler-shifted phase speed $\Omega_-$ for the backward body mode 1 (sbs 1). The phase speed $v_-$ goes upward as $U$ increases without gap appearance. The Doppler-shifted phase speed $\Omega_-$ shows its change of direction at $U/v_{si}\approx0.9$.
\begin{figure}
\includegraphics[width=0.5\textwidth]{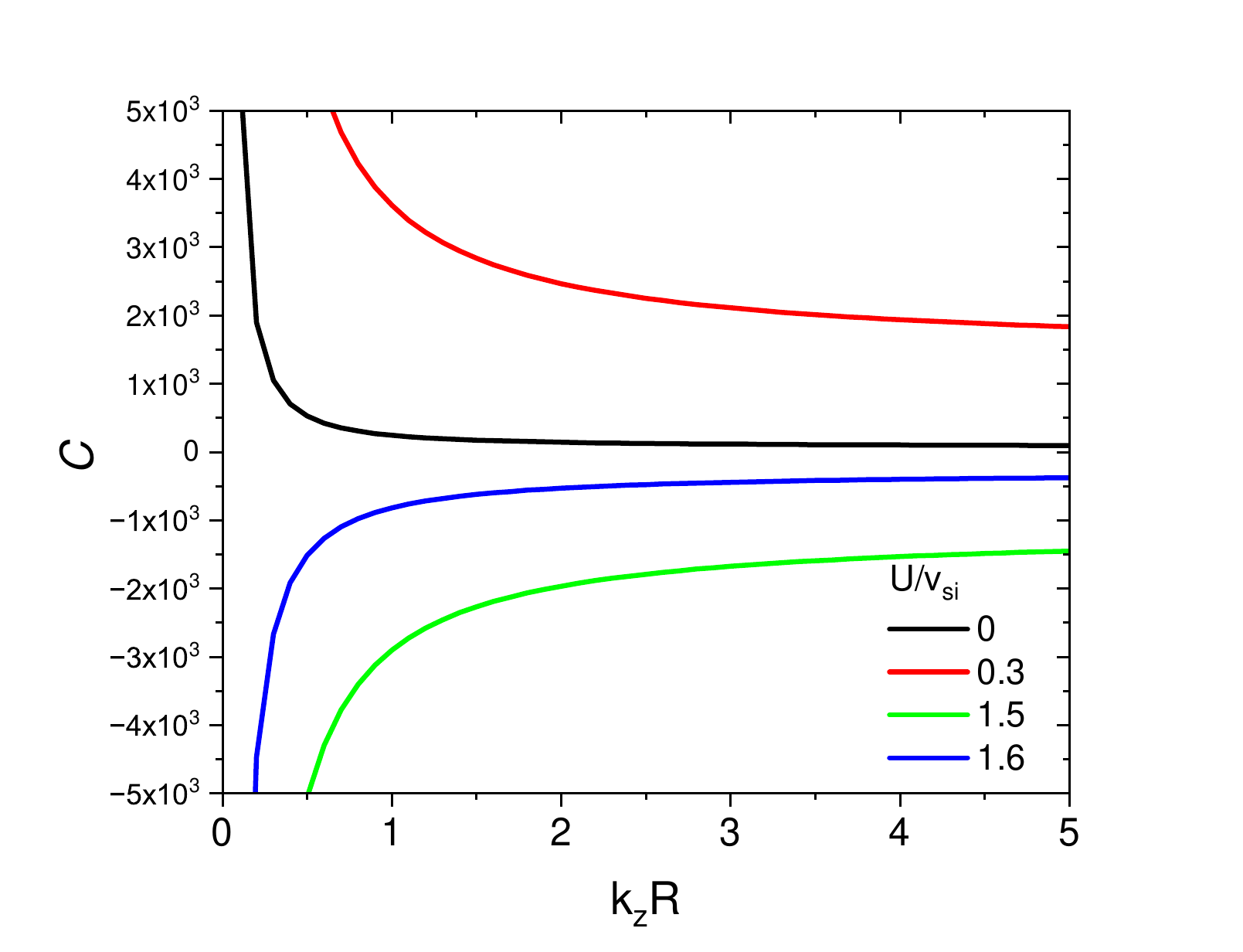}
\caption{\label{fig:f4} The characteristic function $\mathcal{C}$  of the slow surface sausage (sss) mode vs. $k_zR$. }
\end{figure}

\begin{figure}
\includegraphics[width=0.5\textwidth]{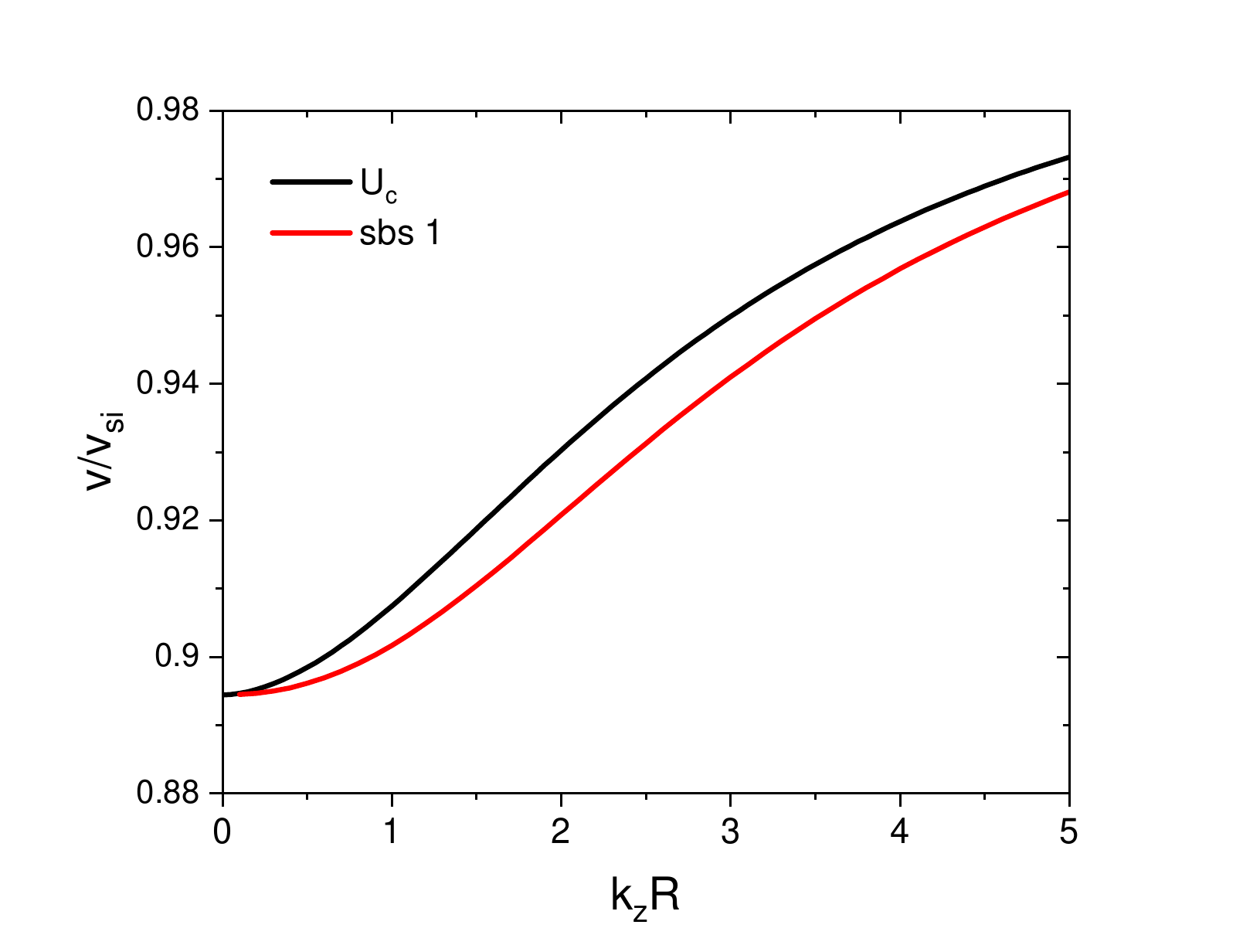}
\caption{\label{fig:f5} The critical speed $U_c$ and dispersion curve of the slow body sausage (sbs) mode 1 vs. $k_zR$. }
\end{figure}

\begin{figure}
\includegraphics[width=0.42\textwidth]{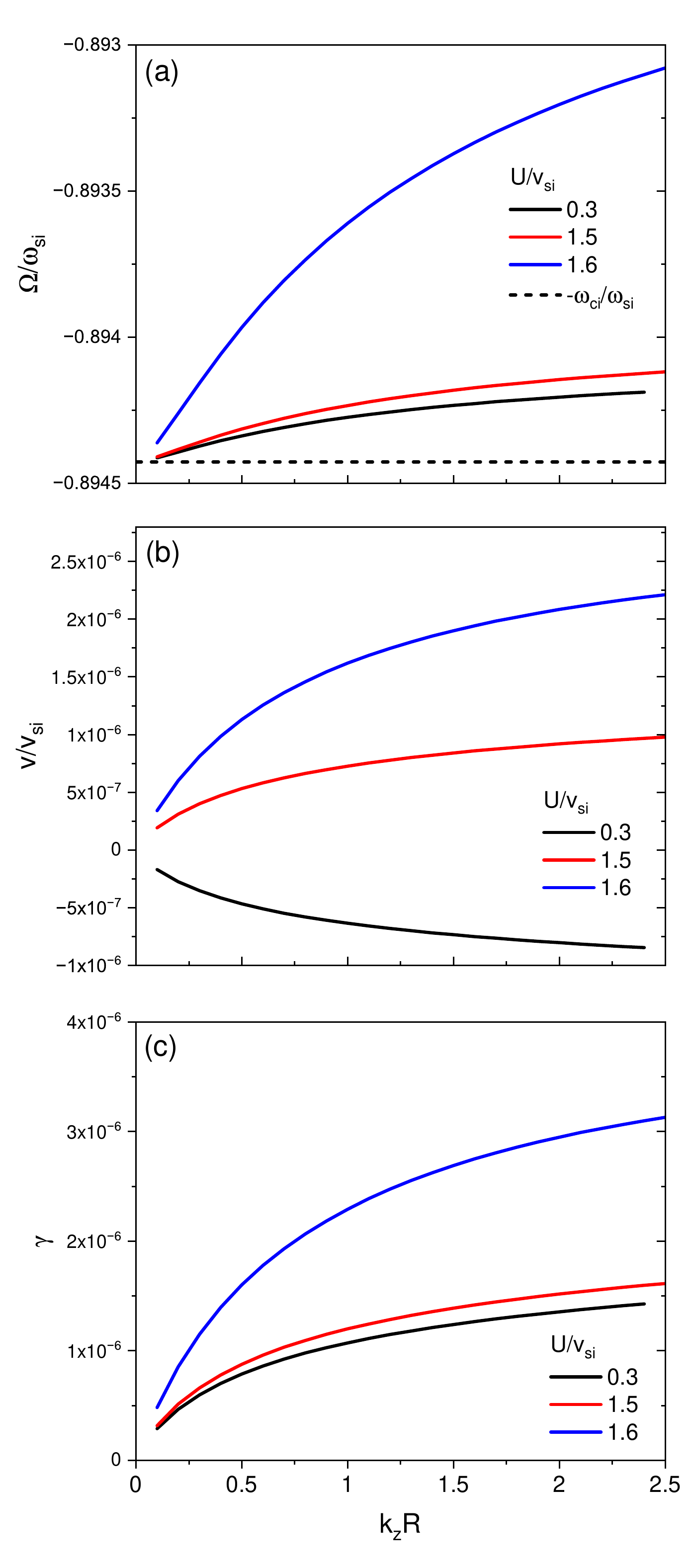}
\caption{\label{fig:f6} (a) Doppler shifted real and (b) imaginary parts of $v_-$, and (c) the ratio $\gamma$ (=Im($v_-$)/Re($v_-$)) of the backward slow surface sausage mode vs. $k_zR$ where $\tilde{\mu}=\tilde{\zeta}=10^{-4}$. .}
\end{figure}

\subsection{Characteristic function $\mathcal{C}$ and critical speed $U_c$}\label{sec:3-2}
In Fig.~\ref{fig:f4} we present the characteristic function $\mathcal{C}$ vs. $k_zR$ for the slow surface sausage mode, using Eqs.~(\ref{eq:28}) and (\ref{eq:29}). This mode becomes NEW when the flow shear is above the gap regime. On the other hand, it is not possible to obtain the critical speed due to the gap appearance.

For the body modes, the situation is opposite.  The characteristic function $\mathcal{C}$ using dispersion function (\ref{eq:29}) is incorrect and of which a suitable form was not found due to its singular behavior in the concerned range of $v_p$. Instead, we obtain the critical speed $U_c$ from the dispersion relation (\ref{eq:22}) with the condition $\omega=0$, as explained in Sec.~\ref{sec:2-4}. Fig.~\ref{fig:f5} gives $U_c$ and $v_-$ vs. $k_zR$ for the slow body mode 1 (sbs 1).
It reveals that for the DI (NEWI) to occur, the flow shear has to be at least higher than the phase speed.
If $U_c$ is fixed, the corresponding $k_z$ is also fixed. By denoting this value $k_c$, there will be a sign change of the phase speed $v_-$ in both real and imaginary parts when $k_z$ goes through $k_c$. When $U/v_{si}=0.9$ $k_cR\approx0.6$.

\begin{figure}
\includegraphics[height=0.62\textheight,width=0.49\textwidth]{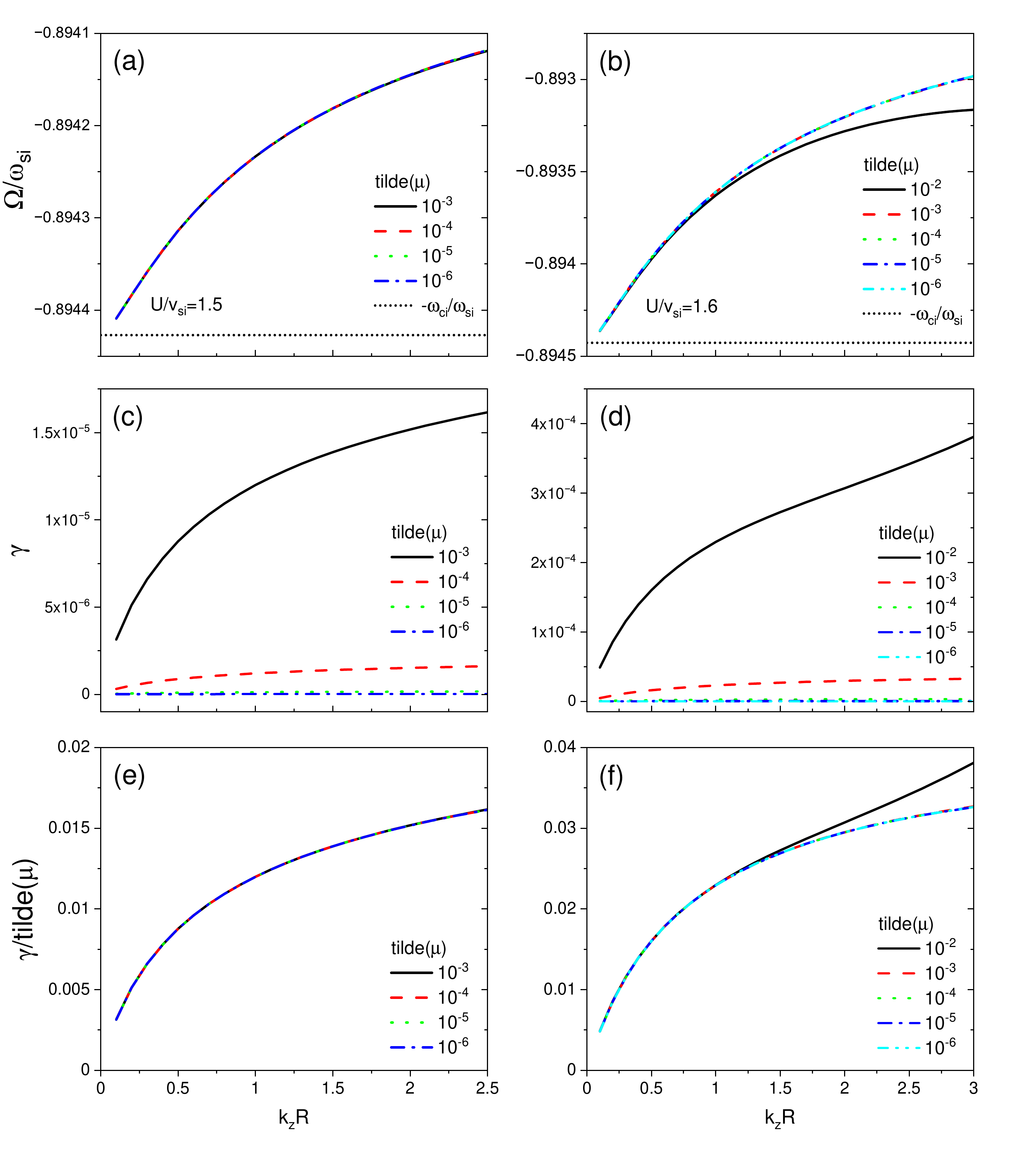}
\caption{\label{fig:f7} Slow surface sausage mode. (a), (b): Re($\Omega_-$) vs. $k_zR$, (c), (d): $\gamma$ vs. $k_zR$, (e), (f): $\gamma/\tilde{\mu}$ vs. $k_zR$ where $\tilde{\mu}=\tilde{\zeta}$. The left and right columns show the $\tilde{\mu}$ dependent behavior for $U/v_{si}$=1.5 and 1.6, respectively. }
\end{figure}

\begin{figure}
\includegraphics[width=0.42\textwidth]{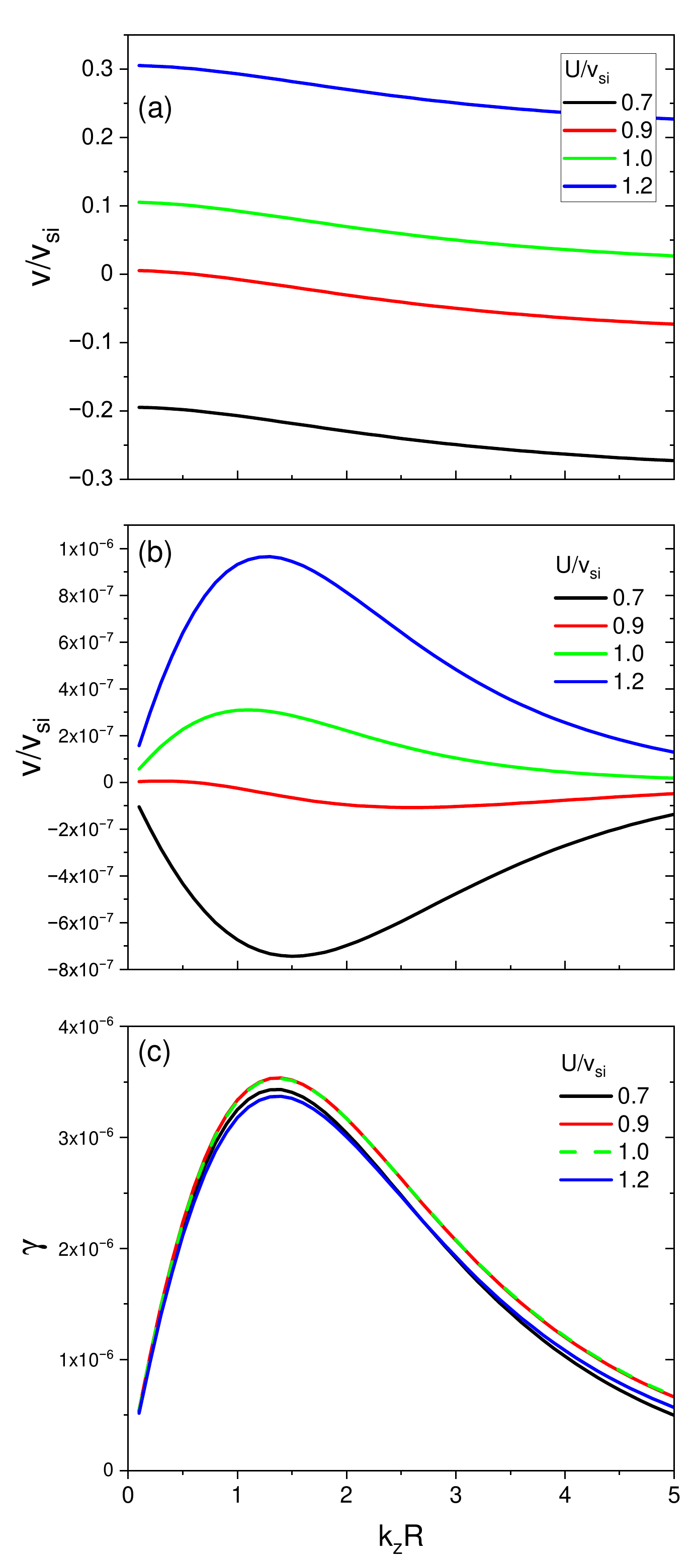}
\caption{\label{fig:f8} (a) Real and (b) imaginary parts of $v_-$, and (c) their ratio $\gamma$ (=Im($v_-$)/Re($v_-$)) for the backward slow body sausage mode 1 vs. $k_zR$ where $\tilde{\mu}=\tilde{\zeta}=10^{-4}$. The real and imaginary parts of the phase speeds change their sign at $k_zR\approx0.6$ when $U/v_{si}=0.9$.  }
\end{figure}

\begin{figure}
\includegraphics[height=0.62\textheight,width=0.49\textwidth]{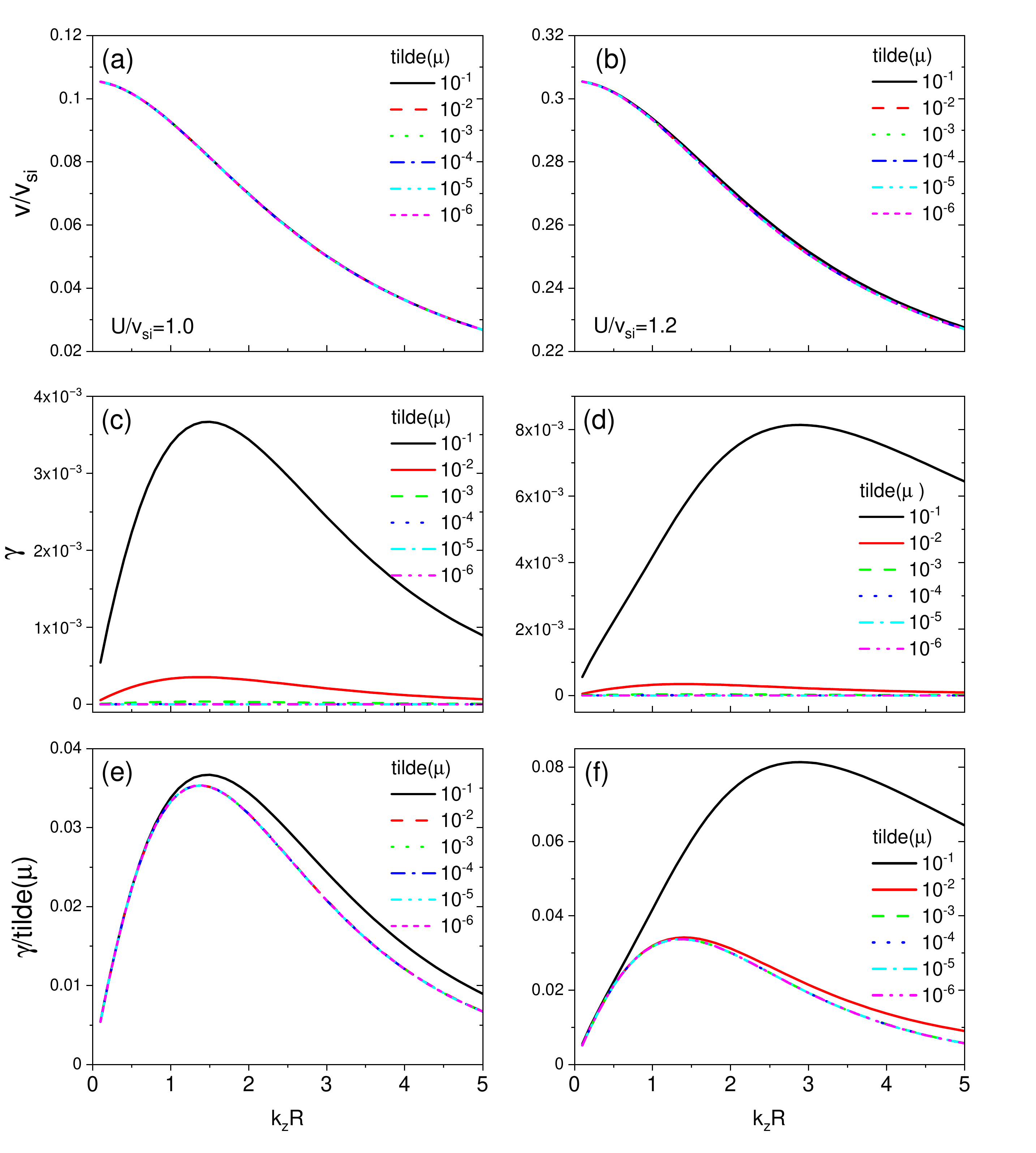}
\caption{\label{fig:f9} Slow body sausage mode 1. (a), (b): Re($v_-$) vs. $k_zR$, (c), (d): $\gamma$ vs. $k_zR$, (e), (f): $\gamma/\tilde{\mu}$ vs. $k_zR$ where $\tilde{\mu}=\tilde{\zeta}$. The left and right columns show the $\tilde{\mu}$ dependent behavior for $U/v_{si}$=1.0 and 1.2, respectively. }
\end{figure}

\subsection{Dispersion curves with shear flow and viscosity}\label{sec:3-3}
In this subsection, we solve the dispersion relation (\ref{eq:36}), considering both flow and viscosity. We first consider the slow surface sausage mode.
In Fig.~\ref{fig:f6}, we plot (a) Doppler-shifted real part and (b) imaginary part of the phase speed $v_-$ and (c) their ratio, $\gamma$ (=Im($v_-$)/Re($v_-$)), as a function of $k_zR$ with varying $U/v_{si}$ where $\tilde{\mu}(=\mu/(\rho_ev_{se}R))=\tilde{\zeta}(=\zeta/(\rho_ev_{se}R))=10^{-4}$. Fig.~\ref{fig:f6} (a) and (b) show that both real and imaginary parts of phase speed $v_-$ shift upward as $U$ increases. Fig.~\ref{fig:f6} (c) indicates that $\gamma$ tends to increase with the increment of $U$ and $k_z$ in the negative energy wave regime ($\mathcal{C}<0$). The same trend was previously shown for the incompressible plasma (Yu \& Nakariakov~\citeyear{Yu2020}). The order of magnitude is $\sim-6$ ($\gamma/\tilde{\mu}\sim10^{-2}$).

It may be more useful to use $\gamma/\tilde{\mu}$ for estimating the strength of the instability. Yu \& Nakariakov (\citeyear{Yu2020}) showed that $\gamma$ of the surface mode in an incompressible plasma is a linear function of $\tilde{\mu}$ ($\tilde{\nu}_e$ therein). Its order of magnitude is $\sim 1$ for underdense and $\sim -1$ for overdense plasmas.
This implies that the strength of dissipative instability in the compressible plasma is weaker than in the incompressible plasma.

We examine  whether the linear relation between $\gamma$ and $\tilde{\mu}$ obtained in an incompressible plasma (Yu \& Nakariakov \citeyear{Yu2020}) is valid or not for the compressional plasma. In Fig.~\ref{fig:f7} we compare two situations, $U/v_{si}=1.5$ (left column) and $U/v_{si}=1.6$ (right column), with variation of $\tilde{\mu}(=\tilde{\zeta})$. In the first row, (a) and (b), the effect of $\tilde{\mu}$ on Doppler-shifted phase speed is shown to be weak for small $\tilde{\mu}$ while the middle row, (c) and (d), reveals that $\gamma$ is in proportion to $\tilde{\mu}$ and $U$. From the bottom row, (e) and (f), we know that the linear relation is valid for small $\tilde{\mu}$ and $k_zR$ and for low $U$. When $U$ is sufficiently high and $\tilde{\mu}$ is relatively large, one may anticipate that $\gamma$ deviates from the linear trend as $k_zR$ increases.

A similar behavior is obtained for the slow body mode (sbs) 1.
In Fig.~\ref{fig:f8}, we plot (a) real and (b) imaginary parts of the phase speed $v_-$ and (c) their ratio $\gamma$ (=Im($v_-$)/Re($v_-$)) vs. $k_zR$ for slow body mode 1, with varying $U/v_{si}$ where $\tilde{\mu}(=\tilde{\zeta})=10^{-4}$.
Both parts of the phase speed shifts upward as $U$ increases and becomes positive when crossing the critical speed. The value of $\gamma$ is always positive except at $k_z=k_c$ and its order of magnitude is about $-6$. Fig.~\ref{fig:f8} (c) presents that $\gamma/\tilde{\mu}\approx10^{-2}$, similar to the slow surface sausage mode. It has a peak at $k_zR\approx1.4$, contrary to the slow surface mode. The axial wavelength at the peak is comparable to the diameter of the flux tube.

In Fig.~\ref{fig:f9} we display two flow cases for the slow body mode 1, $U/v_{si}=1.0$ (left column) and $U/v_{si}=1.2$ (right column), by varying $\tilde{\mu}(=\tilde{\zeta})$.
The top row, (a) and (b), shows the variation of Re($v_-$) vs. $k_zR$. Its dependence on $\tilde{\mu}$ is tiny, but tends to increases as $U$ increases.
The middle row, (c) and (d), describes the growth rate $\gamma$ vs. $k_zR$ while the bottom row, (e) and (f), represents $\gamma/\tilde{\mu}$ vs. $k_zR$. In two cases, the behavior of $\gamma$ looks similar, but its normalized value $\gamma/\tilde{\mu}$ is different. The linear trend is valid only for small $\tilde{\mu}$. After $\tilde{\mu}$ is over a certain point it turns into a nonlinear regime and the peak at $k_zR\approx1.4$ shifts to the right. The threshold of $\tilde{\mu}$ for the deviation to the nonlinear trend becomes smaller as $U$ increases.

\begin{figure}
\includegraphics[height=0.42\textheight,width=0.49\textwidth]{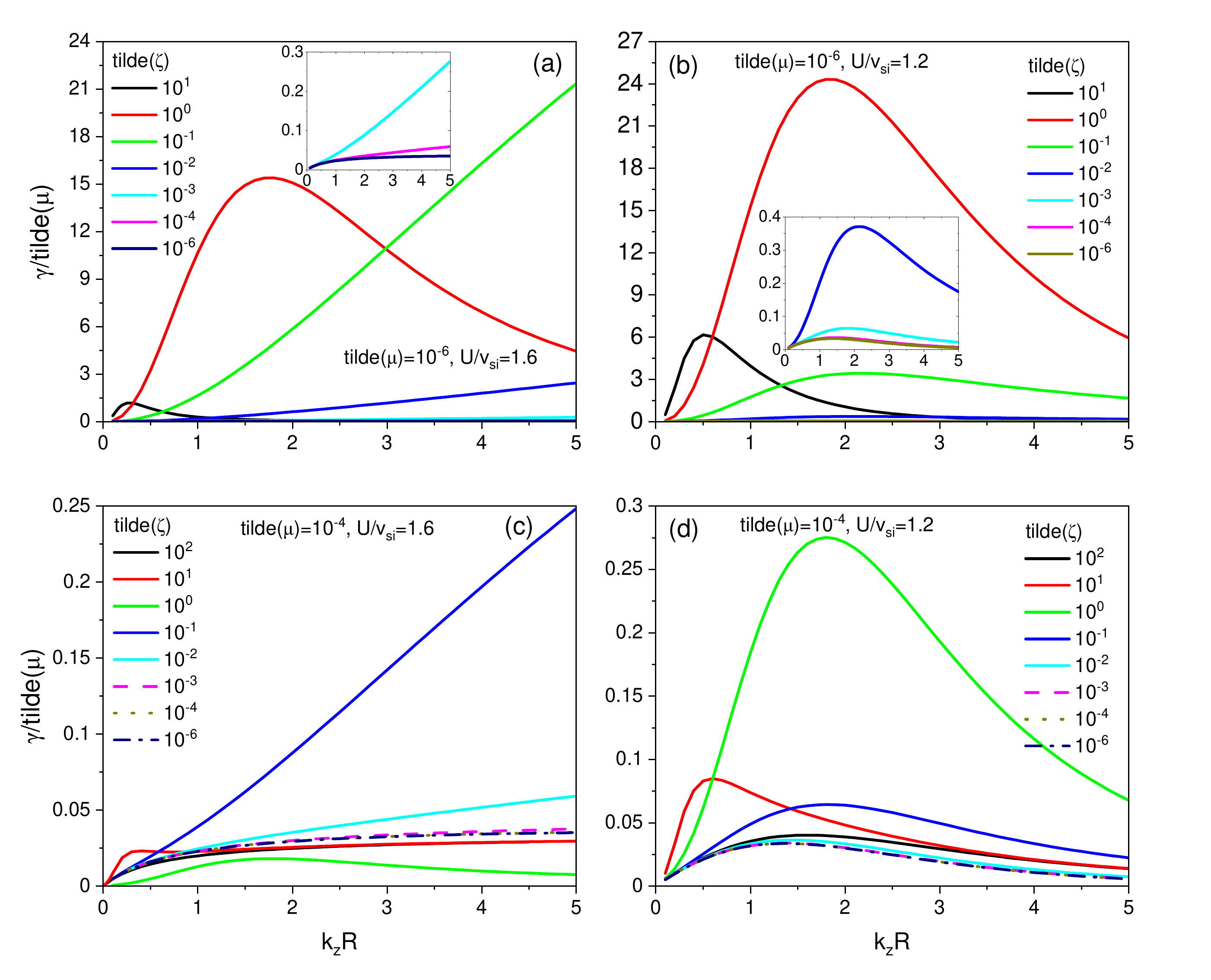}
\caption{\label{fig:f10} The dependence of $\gamma/\tilde{\mu}$ on $\tilde{\zeta}$. The left (right) column represents the slow surface mode (slow body mode 1) with $U/v_{si}$=1.6 (1.2). The panels (a) and (b) are for $\tilde{\mu}=10^{-6}$ and (c) and (d) are for $\tilde{\mu}=10^{-4}$. }
\end{figure}

In Fig.~\ref{fig:f10} we show the effect of $\tilde{\zeta}$ on $\gamma$ for $\tilde{\mu}=10^{-6}$ ((a) and (b)) and $\tilde{\mu}=10^{-4}$ ((c) and (d)). The left (right) column denotes the results for the slow surface mode (slow body mode 1) when $U/v_{si}=1.6~(1.2)$.
The $\tilde{\zeta}$-dependent behavior of the $\gamma/\tilde{\mu}$ looks similar for both modes. When the value of $\tilde{\zeta}$ is less than
that of $\tilde{\mu}$ its effect on $\gamma$ is negligible. When its order of magnitude is greater than that of $\tilde{\mu}$, $\gamma/\tilde{\mu}$ gives a noticeable increment. With a further increment of $\tilde{\zeta}$, $\gamma/\tilde{\mu}$ reaches a maximum in the plot range of $k_zR$, which depends on the value of $\tilde{\mu}$. It is found that when the value of $\tilde{\zeta}$ is sufficiently large a new peak appears for the surface mode. Its position shifts to the left (to smaller $k_zR$) as $\tilde{\zeta}$ increases. On the other hand, for the body mode, the peak position shifts, at first, to the right and then to the left as $\tilde{\zeta}$ increases (panel (b)). When $\tilde{\mu}$ is further increases, the motion of the peak is more complicate. Panel (d) shows that the peak has zigzag motion. Comparing the top and bottom rows, one may notice that the effect of $\tilde{\zeta}$ on the increment of $\gamma/\tilde{\mu}$ becomes weaker as $\tilde{\mu}$ increases.
When $\tilde{\mu}=10^{-2}$, the variation range of $\gamma/\tilde{\mu}$ is tiny and $\gamma/\tilde{\mu}\sim10^{-2}$ (see Fig.~\ref{fig:f7} (f) and Fig.~\ref{fig:f9} (f)). The maximum growth rate is obtained when $\tilde{\zeta}/\tilde{\mu}\sim10^3-10^6$, depending on the wave mode and $\tilde{\mu}$.
 We point out that $\tilde{\zeta}$ rather may give a negative effect on $\gamma$ as shown in panel (c) (the curves for $\tilde{\zeta}\geq10^{0}$).

Cowley's theoretical treatment (\citeyear{Cowley1990}) showed that $\mu\sim4.1\times10^{-4}$g/cms and the order of magnitude of $\zeta/\mu$ can reach 7 in the upper photosphere.
Adopting $\rho_e\approx4.9\times10^{-9}{g/cm^3}$ and $v_{se}\sim10km/s$, we obtain $\tilde{\mu}\sim 8.4\times10^{-4}m/R$. Thus, the more smaller the $R$, the stronger the viscosity. Since $\tilde{\mu}$ has inverse relation with the radius $R$, the waves in the small scale flux tube can be more unstable and more easily excited due to the DI in photospheric regions (see, e.g., Sec. 3.4 in Jess et al.~\citeyear{Jess2023}), contributing to the enhancement of the wave propagations into the upper atmosphere.

\section{Conclusions}
\label{sec:conclusion}
We have studied the DI of sausage modes in a cylindrical circular flux tube in the presence of flow shear and viscosity shear. The equilibrium parameters have different values for the inside and the outside of the flux tube under photospheric condition, having pressure balance at the tube boundary $r=R$. We assume that the equilibrium (backgroud) flow is inside and viscosity outside the flux tube. An analytical formula, Eq~(\ref{eq:36}), for the dispersion relation of the linear sausage modes is derived without approximation.

We first considered the case $\tilde{\mu}=\tilde{\zeta}$.
When $\tilde{\mu}(=\tilde{\zeta})$ is sufficiently small, the growth rate $\gamma$ for the slow surface mode is shown to increase with the increment of $U/v_{si}$ and $k_zR$. This behavior is consistent with that in an incompressible limit, while its magnitude is much smaller. It is also found that the linear relationship between $\gamma$ and $\tilde{\mu}$ previously obtained in the incompressible limit is valid.

For the slow body mode, the behavior of $\gamma$ is slightly different. Its order of magnitude is similar to the slow surface mode, while it has a peak at certain $k_zR$ that depends on $U/v_{si}$ and $\tilde{\mu}$. The DI for the slow body mode is most strong at the axial wavelength comparable to the diameter of the cylinder when $\tilde{\mu}$ is small.

The growth rate of both slow modes grows nonlinearly after $\tilde{\mu}$ crosses a point that depends on $k_zR$ and $U/v_{si}$.

When $\zeta\ne\mu$ the above features change.
The bulk viscosity $\zeta$ plays a more complicate role in the development of the DI.
For small $\tilde{\mu}$, the effect of $\tilde{\zeta}$ on $\gamma$ becomes significant when the order of magnitude of $\tilde{\zeta}$ is 1 or more greater than that of $\tilde{\mu}$ and $\gamma$ reaches a maximum at certain $\tilde{\zeta}$, depending on $\tilde{\mu}$ and $k_zR$. As $\tilde{\mu}$ increases the influence of $\tilde{\zeta}$ on $\gamma$ becomes less important. When $\tilde{\zeta}$ is sufficiently strong, a new peak for the slow surface mode appears and shifts to smaller $k_zR$ with the increment of $\tilde{\zeta}$. The peak for the slow body mode 1 gives more complicate motion depending on the value of $\tilde{\mu}$.

Although the two viscosities are shown to substantially affect the DI, its strength is still weaker than in the incompressible limit, meaning that the compression plays the role of suppressing the DI. Under photospheric condition, the critical speed ranges at the level of sound or cusp speed. To obtain the DI, the critical speed for the DI has to be higher than the phase speed of the slow modes as shown in Fig.~\ref{fig:f5}. It is the phase speed of the wave mode that needs to be considered for estimating the critical velocity.

{Sunspot or pore can be of interest for DI. But, as the smaller radial size the stronger DI, to our view, the detection of DI may be more available for small scale jets (Kotani \& Shibata~\citeyear{Kotani2020}, Muglach~\citeyear{Muglach2021}, Skirvin et al.~\citeyear{Skirvin2023}) or small magnetic loops (Mart{\'\i}nez Gonz{\'a}lez \& Bellot Rubio~\citeyear{MartinezGonzalez2009}, G{\"o}m{\"o}ry et al.~\citeyear{Gomory2010}). }

{The theory for DI developed here does not include ionization effect. For the lower solar atmosphere, especially chromosphere, partial ionization of the plasmas has crucial effect on the wave dynamics (e.g., Leake et al.~\citeyear{Leake2014}, Ballai et al.~\citeyear{Ballai2015}, Ballai et al.~\citeyear{Ballai2017}, Mather et al.~\citeyear{Mather2018}, Alharbi et al.~\citeyear{Alharbi2022}). The strong magnetization affects the viscosity, modifying the viscosity terms used here into more complicated forms (Braginskii~\citeyear{Braginskii1965}). These issues are beyond our scope and will be pursued in a future study.}

As we obtained an analytical solution for the linear MHD sausage waves in a viscous plasma, other applications are possible. For example, the effect of viscosity on the wave damping can be quantitatively analysed and compared with the effect of resistivity on it (see Geeraerts et al~\citeyear{Geeraerts2020}). The DI in the presence of resistivity and flow shear is also an interesting issue to investigate.

\begin{acknowledgments}
The author thanks the referee for valuable comments.
 This research was supported by Basic Science Research Program through the National Research Foundation of Korea(NRF) funded by the Ministry of Education (No. 2021R1I1A1A01045132).
\end{acknowledgments}

\newpage
\appendix
\section{Derivation of the wave equation outside the flux tube}
\label{sec:apped1}
Linearization of Eqs.~(\ref{eq:1})-(\ref{eq:5}) without the background flow leads to
\begin{eqnarray}
\dot{\rho}_1+\nabla\cdot(\rho_0\textbf{{v}}_1)&=&0\label{eq:a1}\\
\rho_0\dot{\textbf{{v}}}_1+\nabla p_1+\frac{\textbf{{B}}_0\times(\nabla\times\textbf{{B}}_1)}{\mu_0}
&=&\mu\nabla^2\textbf{{v}}_1
+\varsigma\nabla(\nabla\cdot\textbf{{v}}_1)~~~~\label{eq:a2}\\
\dot{p}_1+\gamma_0 p_0\nabla\cdot\textbf{{v}}_1&=&0\label{eq:a3}\\
\dot{\textbf{{B}}}_1-\nabla\times(\textbf{{v}}_1\times\textbf{{B}}_0)&=&0,\label{eq:a4}
\end{eqnarray}
where the dot denotes time derivative, $\textbf{{B}}_0=(0,0,B_0)$ , and $\varsigma=\zeta+\mu/3$. Here the subscript 0 and 1 mean zeroth and first order quantities, respectively.
We start by using the notion $\triangle$ for the divergence of $\textbf{{v}}_1=(v_{r},v_{\phi},v_{z})$ (e.g., Geeraerts et al.~\citeyear{Geeraerts2020}) as
\begin{eqnarray}
  \triangle=\nabla\cdot\textbf{{v}}_1=\psi(r)\exp[-i(\omega t-k_zz)].\label{eq:a5}
\end{eqnarray}
Eq.~(\ref{eq:a3}) can be written as with the help of Eq.~(\ref{eq:a1})
\begin{eqnarray}
  \dot{{p}}_1=-\gamma_0\rho_0\triangle=v_s^2\dot{\rho}_1\label{eq:a6}
\end{eqnarray}
where the sound speed is $v_s=\sqrt{\gamma_0 p_0/\rho_0}$.
From Eq.~(\ref{eq:a4}) the perturbed magnetic field $\textbf{{B}}_1=(b_{r},b_{\phi},b_{z})$ can be written as
\begin{eqnarray}
\dot{\textbf{{B}}}_1&=&B_0\nabla\times(v_{1\phi},-v_{1r},0)^T\nonumber\\
&=&B_0\pmatrix{ik_z v_{r} \cr ik_zv_{\phi} \cr
-\frac{1}{r}\frac{\partial (rv_{1r})}{\partial r} \cr }
,\label{eq:a7}
\end{eqnarray}
where the components are
\begin{eqnarray}
b_{r}&=&-\frac{k_zB_0}{\omega} v_{1r},\label{eq:a8}\\
b_{\phi}&=&-\frac{k_zB_0}{\omega}v_{1\phi}=0,\label{eq:a9}\\
b_{z}&=&-\frac{iB_0}{\omega }\frac{\partial(rv_{1r})}{r\partial r}
=-\frac{iB_0}{\omega }\bigg(\frac{v_{1r}}{r}+\frac{\partial v_{1r}}{\partial r}\bigg)
.\label{eq:a10}
\end{eqnarray}

We take time derivative of (\ref{eq:a2}), using Eq.~(\ref{eq:a7}), to obtain
\begin{widetext}
\begin{eqnarray}\label{eq:a11}
\ddot{\textbf{{v}}}_1&=&-\frac{1}{\rho_0}\nabla \dot{p}_1-\frac{1}{\mu_0\rho_0}\textbf{{B}}_0\times(\nabla\times\dot{\textbf{{B}}}_1)
+\frac{\mu}{\rho_0}\nabla^2\dot{\textbf{{v}}}_1
+\frac{\varsigma}{\rho_0}\nabla(\nabla\cdot\dot{\textbf{{v}}}_1)\nonumber\\
&=&v_s^2\nabla \triangle-\frac{1}{\mu_0\rho_0}\textbf{{B}}_0\times[\nabla\times(\nabla\times(\textbf{{v}}_1\times\textbf{{B}}_0))]
+\frac{\varsigma}{\rho_0}\nabla(\dot{\triangle})+\frac{\mu}{\rho_0}\nabla^2\dot{\textbf{{v}}}_1\nonumber\\
&=&v_s^2\nabla \triangle-\frac{1}{\mu_0\rho_0}\textbf{{B}}_0\times[\nabla[B_0(\nabla\times\textbf{{v}}_1)_z]
-\nabla^2(\textbf{{v}}_1\times\textbf{{B}}_0)]
+\frac{\varsigma}{\rho_0}\nabla(\dot{\triangle})+\frac{\mu}{\rho_0}\nabla^2\dot{\textbf{{v}}}_1\nonumber\\
&=&\bigg(v_s^2+\frac{\varsigma}{\rho_0}\frac{\partial}{\partial t}\bigg)\nabla \triangle-\frac{1}{\mu_0\rho_0}\{ B_0\textbf{{B}}_0\times\nabla[(\nabla\times\textbf{{v}}_1)_z]
-\nabla^2[\textbf{{B}}_0\times(\textbf{{v}}_1\times\textbf{{B}}_0)]\}
+\frac{\mu}{\rho_0}\nabla^2\dot{\textbf{{v}}}_1\nonumber\\
&=&\bigg(v_s^2+\frac{\varsigma}{\rho_0}\frac{\partial}{\partial t}\bigg)\nabla \triangle
+v_A^2\pmatrix{(\nabla^2\textbf{{v}}_1)_r \cr (\nabla^2\textbf{{v}}_1)_\phi-\frac{\partial Z}{\partial r} \cr 0 \cr }+\frac{\mu}{\rho_0}\nabla^2\dot{\textbf{{v}}}_1,
\end{eqnarray}
\end{widetext}
where
$Z=(\nabla\times\textbf{{v}}_1)_z=\frac{1}{r}\frac{\partial(rv_{1\phi})}{\partial r}$.

Taking the divergence of Eq.~(\ref{eq:a11}) gives
\begin{eqnarray}\label{eq:a12}
\ddot{\triangle}&=&\bigg(v_s^2-i\omega\frac{\varsigma}{\rho_0}\bigg)\nabla^2 \triangle
+\frac{\mu}{\rho_0}\nabla^2(\nabla\cdot\dot{\textbf{{v}}}_1)
+v_A^2\bigg(\frac{1}{r}\frac{\partial[r(\nabla^2\textbf{{v}}_1)_r]}{\partial r}
\bigg)\nonumber\\
&=&\bigg(v_s^2-i\omega\frac{\varsigma}{\rho_0}\bigg)\nabla^2\triangle
+\frac{\mu}{\rho_0}\nabla^2\dot{\triangle}
+v_A^2\bigg[\nabla^2[\nabla\cdot(\textbf{{v}}_1-\textbf{{v}}_{1z})]\bigg]\nonumber\\
&=&\bar{v}_{{s}}^2\nabla^2\triangle
+v_A^2\bigg[\nabla^2[\nabla\cdot(\textbf{{v}}_1-\textbf{{v}}_{1z})]\bigg]\nonumber\\
&=&(\bar{v}_{{s}}^2+v_A^2)\nabla^2\triangle
-v_A^2\nabla^2\Gamma,
\end{eqnarray}
where $\Gamma=\frac{\partial v_{1z}}{\partial z}$ and $\bar{v}_{{s}}^2=v_s^2-i\omega\frac{\mu+\varsigma}{\rho_0}$.

Next, taking the derivative with respect to $z$ on Eq.~(\ref{eq:a7}) and considering its $z$ component, we obtain for $\Gamma$
\begin{eqnarray}\label{eq:a13}
\ddot{\Gamma}&=&\bar{v}_s^2\frac{\partial^2\triangle}{\partial z^2}
+\frac{\mu}{\rho_0}\frac{\partial^2(\nabla^2\textbf{{v}}_1)_z}{\partial t\partial z}\nonumber\\
&=&\bar{v}_s^2\frac{\partial^2\triangle}{\partial z^2}
+\frac{\mu}{\rho_0}\frac{\partial}{\partial t}(\nabla^2\Gamma).
\end{eqnarray}

Combining Eqs.~(\ref{eq:a12}) and ~(\ref{eq:a13}) we get the governing wave equation for $\triangle$
\begin{eqnarray}\label{eq:a14}
&&\frac{\partial^4\triangle}{\partial t^4}-(\bar{v}_s^2+v_A^2)\nabla^2\ddot{\triangle}\\
&&-\nabla^2\bigg[\frac{\mu}{\rho_0}\frac{\partial^3{\triangle}}{\partial t^3}-\frac{\mu}{\rho_0}(\bar{v}_s^2+v_A^2)\nabla^2\dot{\triangle}
-v_A^2\bar{v}_s^2\frac{\partial^2\triangle}{\partial z^2}\bigg]=0.\nonumber
\end{eqnarray}
This equation reduces to Eq.~(3a) in Edwin \& Roberts (\citeyear{Edwin1983}) when $\mu=\varsigma=0$.

From Eq.~(\ref{eq:a14}) and using the definition of $\triangle$, (\ref{eq:a5}), the perturbed radial velocity $v_r$ can be written in terms of $\triangle$:
\begin{eqnarray}
v_{r}&=&\frac{(v_{s\ast}^2+v_{A\ast}^2)}{\omega^2(\omega^2-v_{A\ast}^2k_z^2)}\bigg[(\omega_{c\ast}^2-\omega^2)
+i\omega^3\frac{\mu}{\rho_0}\frac{v_{A\ast}^2}{v_{A}^2(v_{s\ast}^2+v_{A\ast}^2)}\bigg]
\frac{\partial\triangle}{\partial r}\nonumber\\
&&+i\omega\frac{\mu}{\rho_0}\frac{v_{A\ast}^2(\bar{v}_s^2+v_A^2)}{v_{A}^2\omega^2(\omega^2-v_{A\ast}^2k_z^2)}
\frac{\partial\nabla^2{\triangle}}{\partial r},\label{eq:a15}
\end{eqnarray}
where $v_{s\ast}^2=v_s^2-i\omega\frac{\varsigma}{\rho_0}$, $v_{A\ast}^2=v_A^2-i\omega\frac{\mu}{\rho_0}$, and
$\omega_{c\ast}^2=\frac{v_{A\ast}^2}{(v_{s\ast}^2+v_{A\ast}^2)}k_z^2\bar{v}_s^2$.

From Eq.~(\ref{eq:a10}),  we derive the equation for $b_{z}$ as

\begin{eqnarray}\label{eq:a16}
b_{z}&=&-\frac{iB_0}{\omega }\frac{\partial(rv_{1r})}{r\partial r}
=-\frac{iB_0}{\omega }(\triangle-\Gamma)\nonumber\\
&=&-\frac{iB_0}{\omega}
\bigg\{\triangle-\frac{1}{\omega^2v_{A}^2}\bigg[i\omega^3\frac{\mu}{\rho_0}{\triangle}
+i\omega\frac{\mu}{\rho_0}(\bar{v}_s^2+v_A^2)\nabla^2{\triangle}\nonumber\\
&&+\omega_A^2\bar{v}_s^2\triangle\bigg]\bigg\}\\
&=&-\frac{iB_0}{\omega}
\bigg[1-\frac{\bar{\omega}_s^2}{\omega^2}-i\frac{\mu}{\rho_0\omega v_{A}^2}[\omega^2+
(\bar{v}_s^2+v_A^2)\nabla^2]
\bigg]\triangle\nonumber
.
\end{eqnarray}
Using Eq.~(\ref{eq:a6}) and (\ref{eq:a16}) we obtain the equation for the perturbed total pressure $P$
\begin{eqnarray}\label{eq:a17}
P&=&-\frac{i\rho_0v_A^2}{\omega}
\bigg[1+\frac{v_s^2}{v_A^2}-\frac{\bar{\omega}_s^2}{\omega^2}-i\frac{\mu}{\rho_0\omega v_{A}^2}[\omega^2+
(\bar{v}_s^2+v_A^2)\nabla^2]\bigg]{\triangle}.\nonumber\\
\end{eqnarray}

We note that the equations for $v_r$, $b_z$, and $P$ for the inside of the flux tube can be obtained by setting $\mu=\zeta=0$ and $\omega\rightarrow\Omega$ in the above equations.
\section{Dispersion relation in the presence of shear flow and viscosity}\label{sec:apped2}
Combining Eqs.~(\ref{eq:20}), (\ref{eq:32}), and (\ref{eq:33}), we derive a matrix equation~(\ref{eq:34}), focusing on the body modes,

\begin{eqnarray}
\pmatrix{ X_{1} & X_{2} & X_{3}
\cr X_{4} & X_{5} & X_{6} \cr
X_{7} & X_{8} & X_{9}}
\pmatrix{A_1 \cr A_2 \cr A_3}=\bf{X}\bf{A}=0\label{eq:b1}
,\nonumber
\end{eqnarray}
where
\begin{widetext}
\begin{eqnarray}
X_{1}&=&\frac{\omega}{\Omega}\frac{(\omega_{si}^2+\omega_{Ai}^2)(\omega_{ci}^2-\Omega^2)J_0'}{\Omega^2(\Omega^2-\omega_{Ai}^2)},\label{eq:b2}\\
X_{2}&=&-\bigg[\frac{(\omega_{se\ast}^2+\omega_{Ae\ast}^2)(\omega_{ce\ast}^2-\omega^2)}{\omega^2(\omega^2-\omega_{Ae\ast}^2)}
+i\tilde{\mu}\tilde{k}_z\frac{v_{Ae\ast}^2\omega_{se}\omega}{v_{Ae}^2(\omega^2-\omega_{Ae\ast}^2)}\bigg(1-\frac{(\bar{v}_{se}^2+v_{Ae}^2)(k_+^2+k_z^2)}{\omega^2}\bigg)\bigg]\frac{k_+}{k_i}K_{0+}',\label{eq:b3}\\
X_{3}&=&-\bigg[\frac{(\omega_{se\ast}^2+\omega_{Ae\ast}^2)(\omega_{ce\ast}^2-\omega^2)}{\omega^2(\omega^2-\omega_{Ae\ast}^2)}
+i\tilde{\mu}\tilde{k}_z\frac{v_{Ae\ast}^2\omega_{se}\omega}{v_{Ae}^2(\omega^2-\omega_{Ae\ast}^2)}
\bigg(1-\frac{(\bar{v}_{se}^2+v_{Ae}^2)(k_-^2+k_z^2)}{\omega^2}\bigg)\bigg]\frac{k_-}{k_i}K_{0-}',\label{eq:b4}\\
X_{4}&=&\frac{i\chi\omega v_{Ai}^2}{\Omega v_{Ae}^2}\bigg(1+\frac{v_{si}^2}{v_{Ai}^2}
-\frac{\omega_{si}^2}{\Omega^2}\bigg)J_0,\label{eq:b5}\\
X_{5}&=&\bigg[-i\bigg(1+\frac{v_{se}^2}{v_{Ae}^2}
-\frac{\bar{\omega}_{se}^2}{\omega^2}\bigg)-\tilde{\varsigma}\bigg(\frac{v_{se}R\omega}{v_{Ae}^2}\bigg)
+\tilde{\mu}\bigg(\frac{v_{se}R\omega}{v_{Ae}^2}\bigg)\frac{(\bar{v}_{se}^2+v_{Ae}^2)(k_+^2-k_z^2)}{\omega^2}\bigg]K_{0+}\nonumber\\
&&+\bigg[-2\tilde{\mu}\bigg(\frac{v_{se}R\omega}{v_{Ae}^2}\bigg)
\frac{(\omega_{se\ast}^2+\omega_{Ae\ast}^2)(\omega_{ce\ast}^2-\omega^2)}{\omega^2(\omega^2-\omega_{Ae\ast}^2)}
-i2\tilde{\mu}^2\frac{v_{se}^2\tilde{k}_z^2\omega^2 v_{Ae\ast}^2}{v_{Ae}^4(\omega^2-\omega_{Ae\ast}^2)}
\bigg(1-\frac{(\bar{v}_{se}^2+v_{Ae}^2)(k_+^2+k_z^2)}{\omega^2}\bigg)\bigg]\frac{k_+^2}{k_z^2}K_{0+}'',\label{eq:b6}\\
X_{6}&=&\bigg[-i\bigg(1+\frac{v_{se}^2}{v_{Ae}^2}
-\frac{\bar{\omega}_{se}^2}{\omega^2}\bigg)-\tilde{\varsigma}\bigg(\frac{v_{se}R}{v_{Ae}^2}\bigg)
+\tilde{\mu}\bigg(\frac{v_{se}R\omega}{v_{Ae}^2}\bigg)\frac{(\bar{v}_{se}^2+v_{Ae}^2)(k_-^2-k_z^2)}{\omega^2}
\bigg]K_{0-}\nonumber\\
&&+\bigg[-2\tilde{\mu}\bigg(\frac{v_{se}R\omega}{v_{Ae}^2}\bigg)
\frac{(\omega_{se\ast}^2+\omega_{Ae\ast}^2)(\omega_{ce\ast}^2-\omega^2)}{\omega^2(\omega^2-\omega_{Ae\ast}^2)}
-2i\tilde{\mu}^2\frac{v_{se}^2\tilde{k}_z^2\omega^2 v_{Ae\ast}^2}{v_{Ae}^4(\omega^2-\omega_{Ae\ast}^2)}
\bigg(1-\frac{(\bar{v}_{se}^2+v_{Ae}^2)(k_-^2+k_z^2)}{\omega^2}\bigg)\bigg]\frac{k_-^2}{k_z^2}K_{0-}'',\label{eq:b7}\\
X_{7}&=&\frac{v_{Ai}}{v_{Ae}}\sqrt{\frac{{2v_{Se}^2}+\gamma v_{Ae}^2}{{2v_{Si}^2}+\gamma v_{Ai}^2}}\frac{\omega}{\Omega}
\bigg(1-\frac{{\omega}_{si}^2}{\Omega^2}\bigg)J_0,\label{eq:b8}\\
X_{8}&=&-\bigg[1-\frac{\bar{\omega}_{se}^2}{\omega^2}
-i\tilde{\mu}\bigg(\frac{v_{se}R\omega}{v_{Ae}^2}\bigg)
\bigg(1-\frac{(\bar{v}_{se}^2+v_{Ae}^2)(k_+^2+k_z^2)}{\omega^2 }
\bigg)\bigg]K_{0+},\label{eq:b9}\\
X_{9}&=&-\bigg[1-\frac{\bar{\omega}_{se}^2}{\omega^2}-
i\tilde{\mu}\bigg(\frac{v_{se}R\omega}{v_{Ae}^2}\bigg)\bigg(1-\frac{(\bar{v}_{se}^2+v_{Ae}^2)(k_-^2+k_z^2)}{\omega^2}
\bigg)\bigg]K_{0-}.\label{eq:b10}
\end{eqnarray}
\end{widetext}
The function $K_{0\pm}$ denotes $K_0(k_\pm R)$, $\chi=\rho_i/\rho_e$, $\tilde{\mu}(\tilde{\varsigma})=\mu(\varsigma)/\rho_ev_{se}R$, and $\tilde{k}_z=k_zR$. For the surface modes $J_0$ changes to $I_0$.

\end{document}